\documentclass[fleqn,twoside]{article}
\usepackage{espcrc2}
\usepackage{amssymb}

\usepackage{graphicx}
\usepackage{epsfig}
\newcommand{\beq}{\begin{equation}}
\newcommand{\eeq}{\end{equation}}
\newcommand{\beqn}{\begin{eqnarray}}
\newcommand{\eeqn}{\end{eqnarray}}
\newcommand{\bea}[1]{\beq\begin{array}{#1}}
\newcommand{\eea}{\end{array}\eeq}
\newcommand{\eq}[1]{(\ref{#1})}
\newcommand{\dD}{{\cal D}}
\newcommand{\cD}{{\cal D}}

\newcommand{\LL}{{I\!\! L}}

\newcommand{\intinf}{\int\nolimits^\infty_{-\infty}}

\newcommand{\dd}{{\mathrm d}}
\newcommand{\Z}{Z\!\!\! Z}

\newcommand{\diff}{\partial}

\newcommand{\cZ}{{\cal Z}}
\newcommand{\cS}{{\cal S}}

\title{
\thispagestyle{empty}
\vspace{-25mm}
\rightline{\small KANAZAWA-03-18~~~~~}
\rightline{\small ITEP-LAT-2003-13~~~~~}
\vspace{10mm}
Blocked lattice monopoles in quenched SU(2) QCD \newpage
and dual superconductor model\thanks{Presented by M.~Ch. at Lattice'03.}
}

\author{M. N. Chernodub\address{ITEP, B.Cheremushkinskaya 25, Moscow,
117259, Russia}${}^{\mathrm{,b}}$\thanks{M.N.Ch. is supported by JSPS Fellowship P01023.},
Katsuya~Ishiguro\address{Institute for Theoretical Physics, Kanazawa University,
Kanazawa 920-1192, Japan}
and Tsuneo Suzuki${}^{\mathrm{a}}$\thanks{T.S. is partially
supported by JSPS Grant-in-Aid for Scientific Research on Priority Areas
No.13135210 and (B) No.15340073. This work is also supported by the
Supercomputer Project of the Institute of Physical and Chemical
Research (RIKEN).}
}
\begin{document}

\begin{abstract}
We study the action of the lattice monopoles in
quenched SU(2) QCD in the Maximal Abelian projection. We relate the lattice
action of the monopole currents to the monopole degrees of freedom
of the continuum dual superconductor model and obtain the value of the monopole condensate.
\end{abstract}

\maketitle

The key feature of the dual superconductor mechanism~\cite{DualSuperconductor} of
the color confinement in non--Abelian gauge theories is the Abelian monopole
condensation (for a review, see, $e.g.$, Ref.~\cite{Reviews}).
The mo\-no\-po\-le condensate is formed in the low temperature (confinement)
phase and it disappears in the high temperature (deconfinement) phase~\cite{MonopoleCondensation}.
The monopoles provide a dominant contribution to the tension of the fundamental
chromoelectric string~\cite{AbelianDominance}.

There were various attempts to determine the lagrangian of the dual superconductor and the values of its
couplings~\cite{ref:suzuki:maedan,various:attempts,singh,string:profile,ref:koma:suzuki}.
In these approaches the
solutions of the classical equations of motion of the dual model
were related to the quantum observables in the SU(2) gauge theory.
Our aim is to determine the lagrangian of the dual model in a quantum way.

To this end we compare the {\it lattice} monopole model (obtained numerically) with the
{\it continuum} dual superconductor model using the analytical approach of blocking of the continuum
variables to the lattice~\cite{BlockingOfFields}. This kind of the blocking is ideologically similar
to the lattice blockspin transformation of the monopole action introduced in Ref.~\cite{ref:fujimoto}.
The obtained action of the lattice monopole model depends on the
parameters of the continuum model. Thus, the comparison of the analytical and numerical results
allows us to fix the parameters of the continuum model. In this paper we determine
the monopole condensate in the quenched SU(2) QCD in the Maximal Abelian (MA) gauge.

In order to construct
the lattice monopole action we start from the continuum dual Ginzburg--Landau (DGL):
\beqn
\cZ & = & \int\hspace{-3.3mm}\Sigma \dD k \int \dD B \, \exp\Bigl\{
- \int \dd^4 x\, \Bigl[\frac{1}{4 g^2} F^2_{\mu\nu} \nonumber\\
& & + i k_\mu(x) B_\mu(x)\Bigr] - S_{int}(k)\Bigr\}\,,
\label{eq:Zmon}
\eeqn
where $F_{\mu\nu} = \partial_\mu B_\nu - \partial_\nu B_\mu$ is the field stress tensor of
the dual gauge field $B_\mu$, and $S_{int}(k)$ is the action of the closed monopole currents
$k_\mu$. The integration is carried out over the dual gauge fields and over all possible monopole
trajectories.

The integration over the monopoles gives~\cite{ref:suzuki:maedan}:
\beqn
\cZ & = & \int \dD \Phi \int \dD B \, \exp\Bigl\{
- \int \dd^4 x \, \Bigl[\frac{1}{4 g^2} F^2_{\mu\nu} \nonumber \\
& & + \frac{1}{2} |(\partial_\mu
+ i B_\mu)\Phi|^2 + V(\Phi)\Bigr\}\,,
\label{eq:ZAHM}
\eeqn
where $\Phi$ is the complex monopole field. The self--interactions of the monopole trajectories
described by the action $S_{int}$ in Eq.~\eq{eq:Zmon} lead to the self--interaction of the
monopole field $\Phi$ described by the potential term $V(\Phi)$ in Eq.~\eq{eq:ZAHM}.

Next, we embed the hypercubic lattice with the spacing $b$ into the continuum space.
The $3D$ cube $C_{s,\mu}$ is defined by relations
$\{b (s_\nu - 1/2) \leq x_\nu \leq b (s_\nu + 1/2)$ for all $\nu\neq\mu$ and
$x_\mu = b s_\mu$. Here $s_\nu$ is the dimensionless lattice coordinate of the
cube $C_{s,\mu}$ and $x_\nu$ is the continuum coordinate.

The magnetic charge $K_C$ inside the lattice cube $C_{s,\mu}$ is equal to the total charge
of the continuum monopoles, $k$, passing through the cube. Geometrically,
$K_C$ is given by the linking number between the
cube and the monopole trajectory:
$$
 K_C(k) = \frac{1}{2} \int_k \!\dd x_\mu \int_\Sigma \!
 \dd^2 y_{\nu\alpha} \, \epsilon_{\mu \nu \alpha \beta}
 \diff_{\beta} \cD^{(4)}(x - y)\,.
$$
Here $\Sigma = \partial C$ is the boundary of the cube $C$.

To rewrite the dual superconductor model~\eq{eq:ZAHM} in terms of the lattice currents $K_C$
we insert the unity,
$1 = \sum\nolimits_{K_C\in \Z} \, \prod\nolimits_C\delta( K_C - \LL(\partial C,k))$
into the partition function \eq{eq:Zmon}. Representing this unity
as a functional integral over the variable $\theta_C$ and substitute the
result into Eq.~\eq{eq:Zmon}.
Integrating over the currents we get $\cZ =\sum\nolimits_{K_C\in \Z}
\exp\{-S_{mon}(K)\}$, where
\beqn
e^{-S_{mon}(K)} = \intinf \!\!\!\!\! \dD \theta_C
\exp\Bigl\{- {\tilde S}(\theta) + i (\theta, K) \Bigr\}\,.
\label{eq:Smon:lat}
\eeqn
The action of the lattice fields $\theta$ is expressed as
\beqn
& & \hspace{-5mm} e^{-{\tilde S}(\theta)} \!= \!\int \!\!\dD \Phi \int \!\!\dD B \, \exp\Bigl\{
- \!\!\int \!\dd^4 x \Bigl[\frac{1}{4 g^2} F^2_{\mu\nu} \nonumber\\
& & \hspace{-5mm} + \frac{1}{2} \Bigl|\Bigl[ \partial_\mu +
i(B_\mu+{\tilde B}_\mu(\theta)\Bigr]\Phi\Bigr|^2 + V(\Phi)\Bigr]\Bigr\}\,,
\label{eq:Stilde:lat}\\
& & \hspace{-7mm} {\tilde B}_\mu(\theta;x) = \frac{1}{2} \sum\limits_C \int_{\partial C} \!\!\!\!
\dd^2 y_{\alpha\beta} \, \epsilon_{\mu \nu \alpha \beta} \,
\diff_{\nu} \cD^{(4)}(x - y) \, \theta_C \,.
\nonumber
\eeqn

An exact integration over the monopole and dual gauge gluon fields in
Eq.~\eq{eq:Stilde:lat} is impossible. However, in this work
we are interested in the large--$b$ limit in which the monopole action is dominated
by quadratic interactions~\cite{shiba:condensation,chernodub}:
\beqn
S_{mon}(K) = \sum\nolimits_i g_i S_i (k)\,,
\label{eq:monopole:action}
\eeqn
where $S_i \sim k_{\mu}(s) k_{\mu'}(s')$ and $g_i$ are the monopole couplings.
This type of action can be described by one dual gluon exchange. Therefore we disregard
the fluctuations of the monopole field $\Phi$, which lead to the higher--point
interactions in the effective monopole action~\cite{chernodub}.

In the limit $b\to \infty$ the leading contribution to the
monopole action~\eq{eq:Smon:lat} is
$$
S_{\mathrm{mon}}(K) = \sum_{s,s'} \sum_{\alpha,\alpha'}
K_{s,\alpha} \, \cS_{ss',\alpha\alpha'} \, K_{s',\alpha'}\,,
$$
$$
\cS_{ss',\alpha\alpha'} = \frac{2 \pi}{\eta^2 b^2}
\frac{\delta_{\alpha\alpha'} \delta_{s_\alpha,s_\alpha'}}{\Gamma(0, t_{UV} M^2_B\, b^2)}
\cdot \cD^{(3),-1}_\alpha,
$$
where $\cD^{(3)}_\alpha (\vec s_\perp)$ is the three-dimensional Laplacian acting in
a timeslice perpendicular to the direction $\hat \alpha$,
$\Gamma$ is the incomplete gamma function and $t_{UV}$ is an ultraviolet cutoff.

Next we numerically determine the monopole action
in the quenched SU(2) QCD. We simulate the quenched SU(2)
gluodynamics with the Wilson action, $S(U) = - \frac{\beta}{2} \sum_P {\mathrm{Tr}} U_P$.
We fix the MA gauge to extract
the Abelian gauge field $\theta_\mu(s)$ with the help
the projection of the SU(2) link fields $U_\mu(s)$ to the
Abelian gauge fields, $\theta_\mu(s) = \arg U^{11}_\mu(s)$.
The Abelian field strength is decomposed into two parts, $\theta_{\mu\nu}(s) =
\bar{\theta}_{\mu\nu}(s) +2\pi m_{\mu\nu}(s)$. The elementary
monopole currents are determined in a standard way \cite{degrand},
$k_{\mu}(s) = \frac{1}{2}\epsilon_{\mu\nu\rho\sigma}
\partial_{\nu}m_{\rho\sigma}(s+\hat{\mu})$,
where $\partial$ is the forward lattice derivative.

To study the monopole charges at various physical scales we use the
blockspin transformed monopole currents~\cite{ivanenko},
$$
k_{\mu}^{(n)}(s) = \sum_{i,j,l=0}^{n-1}k_{\mu}(n s+(n-1)\hat{\mu}+i\hat{\nu}
     +j\hat{\rho}+l\hat{\sigma})\,.
$$
Applying an inverse Monte-Carlo method~\cite{chernodub,shiba:condensation}
to the monopole configurations we get the effective monopole action.
In our simulations we have used 200 configurations on $48^4$ lattice. The
MA gauge was fixed with the help of the standard iterative procedure.

\begin{figure}[ht]
\vspace{-6mm}
\centerline{\includegraphics[angle=-0,scale=0.4,clip=true]{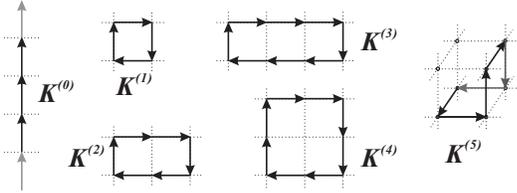}}
\vspace{-5mm}
\caption{The lattice currents used to determine the transverse elements $f_i$ of the monopole action.}
\vspace{-5mm}
\label{fig:closed:curves}
\end{figure}

Note that the shift of the quadratic operator $\cS \to \cS + \alpha \partial \partial'$
(with arbitrary $\alpha$) does not change the monopole action due to
the closeness of the monopole currents. Thus only the transverse part of the
operator $\cS$ has a sense. We evaluate this part
calculating the monopole action on a set of closed trajectories, $K^{(i)}$,
\beqn
f_i(b) \equiv \frac{S(K^{(i)})}{|K^{(i)}|}
= \frac{2 \pi \, d_i}{\eta^2 b^2\, \Gamma(0, b^2 M^2_B t_{UV})}\,,
\label{eq:theor}
\eeqn
where $|K^{(i)}|$ is the length of the trajectory $K^{(i)}$ and $d_i$ are certain numbers
depending on the lattice size.
We consider six types of the trajectories $K^{(i)}$ shown in Figure~\ref{fig:closed:curves}.
\begin{table}
\begin{center}
\begin{tabular}{|c|c|c|c|}
\hline
coupling & $\eta \slash \sqrt{\sigma}$ &
coupling & $\eta \slash \sqrt{\sigma}$ \\

\hline
$f_0$ &  0.521(25) &  $f_1$ &  0.577(41) \\
$f_2$ &  0.565(34) &  $f_3$ &  0.544(32)   \\
$f_4$ &  0.554(28) &  $f_5$ &  0.591(38) \\
\hline
\multicolumn{4}{|c|}{{\bf average:}  {$\mathbf{\eta  = 0.552(13) \, \sqrt{\sigma}}$}}
\\
\hline
\end{tabular}
\end{center}
\caption{The monopole condensate $\eta$ from the fits.}
\vspace{-7mm}
\label{tbl:fitting:parameters}
\end{table}

We used $n=6$ extended mo\-no\-po\-les to
fit of the transverse couplings by functions~\eq{eq:theor}.
\begin{figure}[htb]
\vspace{-5mm}
\centerline{\includegraphics[angle=-00,scale=0.25,clip=true]{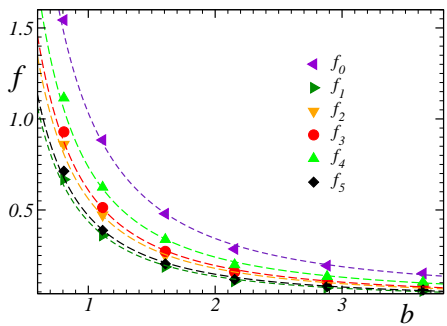}}
\vspace{-8mm}
\caption{The fits of the
monopole couplings.}
\label{fig:alpha:beta}
\end{figure}
The fits are shown in Figure~\ref{fig:alpha:beta} and
the best fit values of the condensate obtained from the independent fits of couplings $f_i$
are presented in Table~\ref{tbl:fitting:parameters}. These values coincide with each other indicating
self--consistency of our approach.

Averaging of the results of the six independent fits and taking into account systematic errors
we get the value of the monopole condensate, $\eta = 243(40)$~MeV. This result is in a {\it quantitative}
agreement with the value~\cite{string:profile}, $\eta=194(19)$~MeV,
obtained by a completely different method. We conclude that the blocking
from continuum is a powerful tool to get the couplings of the DGL model. We are going to
get other parameters of the dual model by this method in
future~\cite{suzuki:chernodub:in:prep}.

\end{document}